# Twist driven deep-ultraviolet-wavelength exciton funnel effect in bilayer boron nitride

Linghan Zhu,[1] Yizhou Wang,[1] Li Yang [1,2]

[1]Department of Physics, Washington University in St. Louis, St. Louis, Missouri 63130, USA.

[2]Institute of Materials Science and Engineering, Washington University in St. Louis, St. Louis, Missouri 63130, USA.

ABSTRACT

Realizing direct-bandgap quantum dots working within the deep-ultraviolet frequency is highly desired for electro-optical and biomedical applications while remaining challenging. In this work, we combine the first-principles many-body perturbation theory and effective Hamiltonian approximation to propose the realization of arrays of deep-ultraviolet excitonic quantum dots in twisted bilayer hexagonal boron nitride. The effective quantum confinement of excitons can reach $\sim 400\ meV$ within small twisting angles, which is about four times larger than those observed in twisted semiconducting transitional metal dichalcogenides. Especially because of enhanced electron-hole attraction, those excitons will accumulate via the so-call exciton funnel effect to the direct-bandgap regime, giving the possibility to better luminescence performance and manipulating coherent arrays of deep-ultraviolet quantum dots.

**Introduction.** Quantum dots (QDs), also known as "artificial atoms", have been widely utilized as fundamental building blocks in various applications, including light emitters/detectors[1,2], photovoltaics[3], biomedical sensors[4], and photo-/electro-catalysis[5,6]. Recent advancements in twisted two-dimensional (2D) van der Waals (vdW) bilayers have opened up new possibilities for creating arrays of QDs with uniform sizes[7,8]. The formation of moiré superlattices in these systems, with periodic potential landscapes spanning hundreds or thousands of unit cells, naturally creates quantum wells for carrier confinement. Twisted transition metal dichalcogenides (TMDs) heterostructures have particularly been demonstrated to hold promise in realizing a variety of quantum emitters with high purity[9–12]. However, most intrinsic few-layer TMDs are indirect-bandgap semiconductors, limiting their optical quantum yield and efficiency[13]. To overcome this limitation, heterostructures with a type II band alignment have been employed to create few-layer systems with direct bandgaps[12,14]. Electronically doped TMD moiré heterostructures have been shown to exhibit Mott insulating states and generalized Wigner crystals with atomic-like orbitals[15–18]. Excitonic quantum emitter arrays are proposed in twisted TMD bilayers under electric field[8]. Nonetheless, the moiré confinement potential induced in these TMDs systems is typically less than 100 meV[7–9,19,20], making their optical performance and correlated properties temperature-sensitive, and the increased material complexity and required external fields further limits their broader applications.

Despite significant advancements in twist-driven TMD QDs operating in the infrared/visible frequency, there has been limited proposals to extend their functionality into the deep ultraviolet (DUV) regime, covering wavelengths from 280 nm down to 200 nm. The DUV domain is crucial for broad applications, including polymer curing[21], air-water purification[22], bio-medical instrumentation[23], quantum sensing[24], and germicidal systems[25]. Layered hexagonal boron nitride ($h$-BN) presents itself as an attractive candidate due to its ultra-wide bandgap (UWBG) ($\sim 6.2\ eV$ in the bulk structure)[26–28]. However, intrinsic few-layer and bulk $h$-BN exhibit indirect bandgaps[26], resulting in the quenching of photoluminescence and limitation of their potential for optical applications.

In this work, we employed the first-principles simulations to propose twist-induced DUV QD arrays. We show that by twisting bilayer $h$-BN structures, one can realize QDs with a significant periodic potential confinement of ~400 meV. Importantly, the minimum bandgap within the $h$-BN superlattice is switched to be direct, giving hope to realizing arrays of DUV QDs with an improved quantum yield for optical responses. Going beyond single-particle descriptions, we further incorporate excitonic effects via the many-body perturbation theory (MBPT) and explore an exciton funnel effect, wherein the large moiré exciton binding drives photo-excited electron-hole pairs towards the potential minimum dominated by direct excitons. This exciton funnel effect helps concentrate excitons for energy harvesting and enables coherent optical responses of periodic DUV QDs in an intrinsic moiré superlattice.

**Results and Discussion.** There are two fundamental stacking orders in $h$-BN. Natural bulk $h$-BN takes the AA' stacking, where there is a 180° rotation between adjacent layers[26,29]. Two BN monolayer sheets can also be stacked without rotation, realizing the other AA stacking. A small twist between bilayer $h$-BN gives rise to the moiré superlattices where there is a local variation of interlayer stackings within the supercell. Figures 1(a) and (b) show the moiré superlattices formed by twisting AA and AA' stacked bilayer $h$-BN, respectively. For the twisted AA stacking, two local stackings, AB and BA, are equivalent in the sense that they are related by a spatial inversion. For the twisted AA' stacking, three distinct high-symmetry stackings AA', AB1 and AB2 can be identified. Recent theoretical and experimental studies have demonstrated intrinsic ferroelectricity in twisted AA *h*-BN homobilayers since AB and BA configurations break the inversion symmetry[30–32]. On the other hand, all three stackings of the twisted AA' bilayer keep the inversion symmetry, and no ferroelectricity is expected. In this work, we mainly focus on the twisted bilayer *h*-BN based on the intrinsic AA' stacking. However, the twist-induced direct bandgap and exciton funnel effect proposed in this paper could also be applied to other moiré superlattices.

Lattice reconstruction is considered to be an important issue in the study of twisted vdW materials[33,34]. We have employed LAMMPS[35] to study the local strain introduced by moiré superlattices in twisted bilayer *h*-BN. We adopted the modified Kolmogorov and Crespi (KC) potential[36] to simulate the interlayer interactions and the Tersoff potential[37] for the intralayer interactions. The relative change in local bond length $\delta$ and atomic displacement $\eta$ are defined by

$$\delta = \frac{\Sigma_{i,j} d_{ij} - 3d_{BN}}{3d_{BN}}, \quad (1)$$

$$\eta = \frac{|\Sigma_{i,j} \vec{d_{ij}}|}{3d_{BN}}, \quad (2)$$

where the summations are over the nearest neighbors, and $d$ is the bond length. $d_{BN}$ is the intrinsic bond length. As shown in Figures 1(c) and (d), the local change of bond length and atomic displacement are generally below 0.5% for both metrics in a 0.5° twisted homobilayer *h*-BN. This is expected to affect the electronic structures at the level of ~10 meV[38], which will have little impact on the UWBG *h*-BN. Moreover, previous works have shown that the lattice reconstruction induced strain is inversely proportional to the twist angle[39]. Hence, in the remaining of the paper, we will not consider the strain effect in the interpolation of local band energies for relevant twisting angles between 0.5° to 2°.

*Stacking Dependent Quasiparticle Bandgap:* We begin our discussions of the electronic structures of three high-symmetry stackings in twisted AA' bilayer with the density functional theory (DFT) results. The electronic band structures of AA', AB1, and AB2 stacked bilayer *h*-BN are shown in Figures 2(a)-(c), respectively. The DFT results are represented by the blue dashed lines. The distinct interlayer atomic configurations of different stackings result in substantially different band structures. Intrinsic AA' and AB2 stackings take an indirect bandgap, where the valence band maximum (VBM) is at $K$ and the conduction band minimum (CBM) lies at $M$. In contrast, the AB1 stacking has a direct bandgap at $K$. These findings agree with previous works on the electronic structure of bilayer BN[40,41]. The distinction in band structures can be understood from the atomic arrangements of different stackings. In the AB1 stacking, boron atoms in the top layer are vertically aligned with the boron atoms in the bottom layer. From the projected density of states (PDOS) of the AB1 stacked bilayer BN [see the Supplemental Material[42]], the CBM is mainly contributed by the *p* orbitals of boron atoms. As a result, the strong interlayer interaction and orbital hybridization lead to a large splitting of CBM at $K$, resulting in a direct bandgap. Similarly, in the AB2 stacking, nitride atoms in the top layer lie above the nitride atoms of the bottom layer. As the VBM of bilayer BN is mainly composed of the *p* orbital of Nitride atoms, this renders the splitting of VBM in the AB2 stacking.

Since DFT is known to underestimate bandgaps of semiconductors and insulators, we further performed the GW approximation to obtain the quasiparticle energies. As shown by the red-solid lines in Figures 2(a)-(c), the GW correction does not alter the band alignment but mainly renormalizes the band energies. The direct bandgap in the AB1 stacking is enlarged by $\sim 2.33\ eV$, and the indirect bandgaps in AA' and AB2 stackings are enlarged by $\sim 2.37\ eV$. The self-energy corrections are similar in all three stackings since it is mainly determined by the dimensionality and subsequently reduced electronic screening. The DFT and GW calculated direct and indirect bandgaps are summarized in Table I.

*Twist Driven Continuous Bandgap Variation:* We adopt the effective Hamiltonian to obtain the quasiparticle moiré potential in the twisted bilayer *h*-BN superlattices. The periodic quasiparticle bandgap variation in moiré superlattices can be approximated by a Fourier expansion over the nearest moiré reciprocal lattice vectors[43]

$$V(\boldsymbol{r}) = E(\boldsymbol{r}) = T_0 + 2V_0 \Sigma_{i=1,2,3} \cos(\boldsymbol{b}_i \cdot \boldsymbol{r} \pm \psi), \qquad (3)$$

and it is plotted in Figure 2(d) along the three high-symmetry local stackings. The fitting parameters $(T_0, V_0, \psi)$ from *ab initio* GW calculations for the bandgap are $(6.67\ eV, 41\ meV, 26°)$ for this $C_3$ symmetry system. The overall variation of quasiparticle bandgap within the moiré superlattice is $\sim 430\ meV$, which is significantly larger than those realized in twisted TMDs. Moreover, as demonstrated by the background color in Figure 2 (d), the region around the moiré potential minimum at the AB1 stacking assumes the direct bandgap, which gives hope in realizing high quantum-yield QD arrays in twisted bilayer *h*-BN. We notice that previous works on strained monolayer TMDs[44–46] can realize similar energy barriers under indirect bandgaps. However, by exploiting the natural bandgap variation in twisted homobilayers, we bypassed the requirement of strain engineering of ultra-strength materials, making our approach more feasible to realize in experiments.

Nonetheless, the photo-excited carrier dynamics and, particularly, luminescence are not only decided by the bandgap. Within the single-particle picture, photo-excited carriers in the superlattice tend to move towards the band extrema, as elucidated in Figure 2(e) by the grey arrows for electrons (CBM) and holes (VBM), respectively. Interestingly, although the AB1 stacking has the minimum bandgap, its VBM is lower than that nearby AB2 stacking because of the significant VBM splitting at $K$ point of the AB2 stacking (Figure 2 (c)). As a result, a charge-transfer picture is expected within the single-particle picture, in which excited electrons are accumulated in AB1 stacking while excited holes are accumulated in AB2 stacking.

Such a spatial separation of electrons and holes will diminish optical radiative recombination because of the small overlap between their wavefunctions, which leads to quenched dipole oscillator strength. Fortunately, in the following, we will show that the strong electron-hole interactions and exciton funnel effect in twisted bilayer BN may overcome the spatial separation of electrons and holes and enhance the luminescence intensity.

*Exciton Funnel Effect in h-BN Moir*é *Superlattices:* It is widely recognized that the optical responses of 2D semiconductors and insulators are dictated by excitonic effects, which are dramatically increased due to reduced dielectric screening[13,49]. For this consideration, we have further solved BSE and calculated excitonic effects in the three high-symmetry stackings, the optical absorption spectra of which are shown in Figures 3(a)-(c). Due to the reduced dimensionality and dielectric screening, enhanced electron-hole interactions dominate the optical absorptions spectra. For the AB1 stacking, the lowest-energy exciton state is a direct exciton $X^d$ located at 4.94 eV with a 1.5 eV of electron-hole binding energy. It is mainly formed by optical transitions from holes to electrons around $K$ in the reciprocal space. We also find that, around 500 $meV$ above $X^d$, there is an indirect exciton $X^i$ featuring transitions mainly from $K$ to $M$ with a finite momentum. Because of the requirement of momentum conservation for vertical transitions, the indirect exciton $X^i$ is an optically dark state.

Interestingly, for the indirect-band-gap stackings AA' and AB2, the lowest-energy exciton state is essentially still a direct exciton, $X^d$ originated from transitions around the $K$ point with a zero momentum, as marked in Figure 3 (b) and (c), respectively. This is because the strength of electron-hole interactions is proportional to the electronic structure joint density of states (JDOS)[47]. The excitonic transition matrix elements are a coherent sum of the contributing electron-hole pair transitions, which can also be written as an integral in the energy space[48,49]

$$\langle 0|\hat{v}|i\rangle = \sum_{vck} A^i_{vck} \langle vk|\hat{v}|ck\rangle = \int S_i(\omega)\, d\omega,$$

where $S_i(\omega) = \sum_{vck} A^i_{vck} \langle vk|\hat{v}|ck\rangle \delta(\omega - (E_{ck} - E_{vk}))$ is essentially the electron-hole coupling coefficient modulated JDOS. This also agrees with the picture of the hydrogenic model, in which a larger effective mass contributes to a larger JDOSS and electron-hole binding energy.

As shown in Figures 2 (a) and (c), the CBM at $K$ is doubly degenerate for these two stacking styles while they are split at CBM ($M$ point), which indicates much higher JDOS around $K$ than that at $M$. The enhanced electron-hole interaction induces a large binding energy, which pushes the zero-momentum (direct) exciton state $X^d$ to a lower energy than the indirect exciton state $X^i$. The lower-energy direct exciton in indirect-band-gap bilayer *h*-BN has been discovered in a previous theoretical study[50]. The direct and indirect exciton energies of different stackings are summarized in Table I. The direct $X^d$ exciton binding energy in bilayer BN is large, reaching over 1.5 $eV$ in AB1 and ~1.8 $eV$ in AA' and AB2. Although the JDOS and electron-hole binding energy are smaller in the AB1 stacking than in the other two indirect gap stackings, its quasiparticle bandgap is the smallest, which leads to its lowest-energy bright $X^d$ state among three high-symmetry sites.

The typical real-space wavefunctions of the excitonic state $X^d$ are plotted in the inset of Figure 3(a). Compared with monolayer TMD, in which the typical exciton size is ~$2.6\ nm$[9,51], the exciton radius of bilayer $h$-BN is significantly smaller (~$0.7\ nm$) because of the weaker screening and stronger electron-hole binding. In twisted bilayer $h$-BN, if we take a twist angle of $1°$, the moiré lattice constant is ~$14\ nm$. The radius of the exciton is much smaller than the superlattice, hence we can adopt Eq. (1) and treat the lowest-energy exciton as a composite particle moving within a slowly varying excitonic moiré potential. The moiré potential of exciton energies along the high-symmetry line is plotted in Figure 4(a), and all those excitonic states are direct (zero-momentum). The variation of direct exciton energy within the moiré period is ~$400\ meV$, which is comparable to that reported in strain engineered monolayer MoS$_2$ under a high strain level (~$500\ meV$ at biaxial strain of 5%)[44] and much larger than that in twisted bilayer TMDs (~$100\ meV$)[7,9,52].

Importantly, because the exciton binding energy (~$1.5 - 1.8\ eV$) is much larger than the quasiparticle energy variation (~$300\ meV$), the photoexcited carriers tend to form electron-hole pairs because of the larger energy gain, and their motion is dictated by the energy landscape of excitons, instead of that of single particles shown in Figure 2(e). A schematic plot of the electron-hole interaction modified band edge landscape is shown in Figure 4(b). While the electrons in conduction bands still tend to move towards the overall CBM around the AB1 stacking, the strong exciton binding drags holes in the valence band towards AB1, instead of the overall VBM around AB2. This is essentially the type III exciton funnel effect as proposed previously[44,60], and is a direct result of the overwhelming electron-hole interaction. Therefore, despite the different localizations of VBM and CBM, photoexcited charge carriers (both holes and electrons) will form excitons and accumulate around AB1 stacking in favor of the decreasing exciton energy profile via the exciton funnel effect.

Figure 4(c) summarizes the exciton energy profile in the moiré superlattice. Because of the accumulation of carriers at the lowest-energy exciton state, excitons will be localized around the AB1 high-symmetry sites. The increased local exciton density will further enhance the intensity of coherent photoluminescence. Therefore, by exploiting the periodic quasiparticle bandgap variation and exciton funnel effect in moiré superlattices of bilayer $h$-BN, we can realize arrays of DUV QDs formed by bright exciton states. Finally, it is worth mentioning that the exciton diffusion length reported in 2D materials is typically on the order of ~$1\ \mu m$[53,54], which is much larger than the moiré period discussed in this work. This validates the proposed exciton funnel effect in the moiré superlattices before the excitons recombine.

Experimental realizations of quantum emitter arrays in strain textured monolayer MoS$_2$[55] and WSe$_2$[56] were reported previously by utilizing indented nanopillars, confirming the theoretical predictions of exciton funnel effect in strain engineered continuously varying band gap landscape. Besides, we note that recently, there are experimental efforts in realizing the quantum-dot light emitters in MoS$_2$/WSe$_2$ heterostructures[57], and in hexagonal[58] and orthorhombic[59] boron nitride crystals, via the formation of defect states. Our work yet provides another general and intrinsic approach for achieving excitonic quantum dot arrays in moiré superlattices, which can be verified by photoluminescent measurements on twist-stacked exfoliated monolayer BN.

**Conclusion.** By exploiting the large bandgap variation in twisted bilayer *h*-BN, we propose the realization of bright DUV QDs in the moiré superlattice with the help from exciton funnel effect. Through many-body perturbation theory calculations, we show the large quasiparticle bandgap and exciton energy confinement within the moiré structure are significant ($\sim 400\ meV$). Although the CBM and VBM are located in different high symmetry areas of the moiré superlattice, the large electron-hole binding energy drives both holes and electrons towards the AB1 stacking to form bright direct [55]excitons owing to the exciton funnel effect. The DUV wavelength excitonic QD arrays could have broad applications in energy harvesting and photolithography.

**Computational Details:** The ground-state properties of bilayer *h*-BN are calculated by DFT within the general gradient approximation (GGA) with the Perdew-Burke-Ernzerhof (PBE) exchange-correlation functional[60] as implemented in Quantum ESPRESSO[61]. The vdW interactions are included via the semiempirical Grimme-D3 scheme[62]. The plane-wave energy cutoff is set to 65 Ry under the norm-conserving pseudopotentials. The vacuum level of 18 Å is chosen between adjacent BN bilayers along the out-of-plane direction to avoid spurious interactions. The MBPT simulations are performed with BerkeleyGW[63]. The quasiparticle energies are calculated under the single-shot $G_0W_0$ approximation within the general plasmon pole model[64], where the dielectric matrix energy cutoff is 10 Ry, and over 200 unoccupied bands are utilized for the summation. A coarse k-grid of 18x18x1 is used, which is further interpolated to a fine k-grid of 36x36x1 in order to obtain the electron-hole interaction kernel and solve the Bethe-Salpeter equation (BSE) for the excitonic effects and optical absorption spectra[48].

**Supporting Information:** The projected density of states of AB1 bilayer *h*-BN and the single particle carrier wavefunctions in the moiré superlattices.

**Acknowledgments:**

L.Z. and L.Y. are supported by the National Science Foundation (NSF) grant No. DMR-2124934. Y.W. is supported by the Air Force Office of Scientific Research (AFOSR) grant No. FA9550-20-1-0255. The simulation used Anvil at Purdue University through allocation DMR100005 from the Advanced Cyberinfrastructure Coordination

Ecosystem: Services & Support (ACCESS) program, which is supported by National Science Foundation grants #2138259, #2138286, #2138307, #2137603, and #2138296.

**Table I** DFT/GW bandgaps and exciton energies for three high-symmetry local stackings. The lower-energy values of each stacking are underscored. The global energy minima are in bold font. The energy unit is in eV.

|  | $E_{g,K \to K}^{DFT}$ (direct) | $E_{g,K \to M}^{DFT}$ (indirect) | $E_{g,K \to K}^{GW}$ (direct) | $E_{g,K \to M}^{GW}$ (indirect) | $E_x^d$ (direct) | $E_x^i$ (indirect) |
|---|---|---|---|---|---|---|
| AA' | 4.74 | <u>4.52</u> | 7.19 | <u>6.89</u> | <u>5.30</u> | 5.38 |
| AB1 | **<u>4.13</u>** | 4.51 | **<u>6.46</u>** | 6.90 | **<u>4.94</u>** | 5.45 |
| AB2 | 4.52 | <u>4.29</u> | 6.96 | <u>6.65</u> | <u>5.18</u> | 5.21 |

**Figures:**

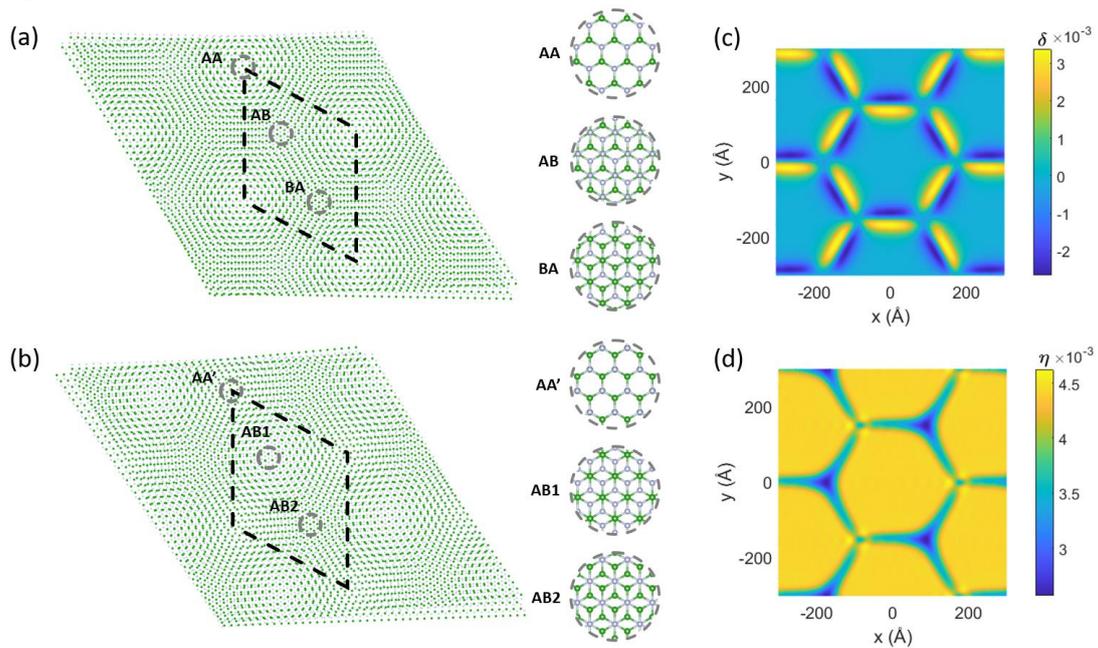

**Figure 1** Moiré superlattices formed by a twist from (a) AA and (b) AA′ stacked bilayer *h*-BN. The top views of the atomic structure of local high-symmetry stackings are shown in the circles. (c) The relative local bond length change $\delta$ and (d) atomic displacement $\eta$ in a fully relaxed 0.5° twisted bilayer *h*-BN.

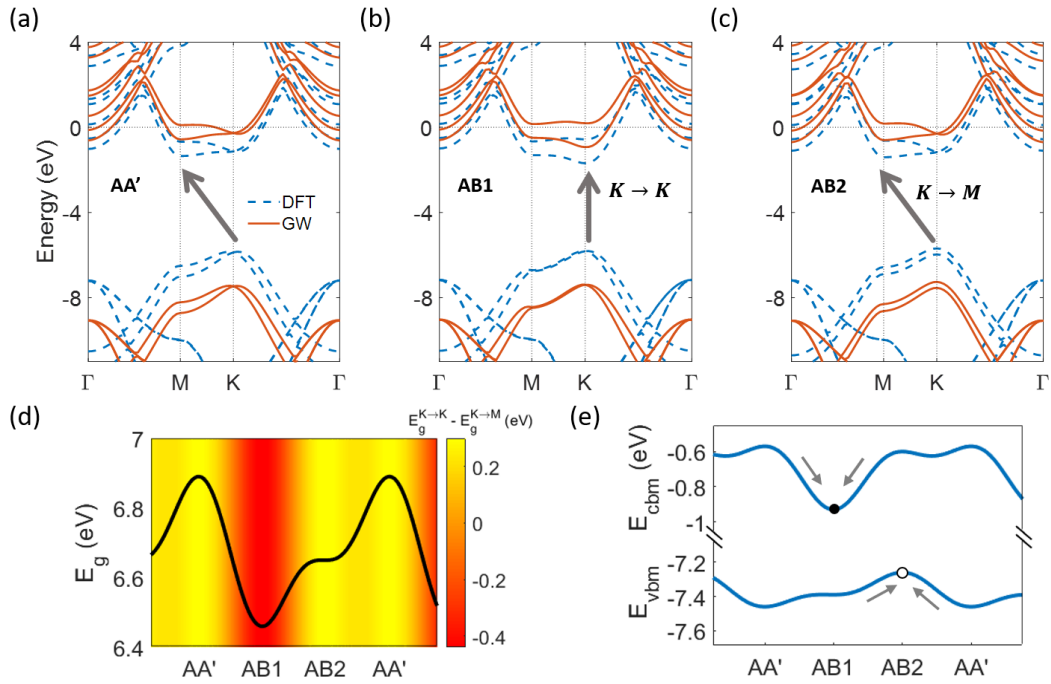

**Figure 2** DFT (dashed blue line) and GW (solid red line) band structures of (a) AA' (b) AB1 (c) AB2 bilayer *h*-BN stackings, respectively. The zero energy is set to the vacuum level. (d) The plane-wave interpolated bandgap landscape in the twisted AA' bilayer *h*-BN moiré superlattice. The background color bar represents the energy difference between the direct gap at $K$ and the indirect gap from $K$ to $M$. Notice that the deep color represents the region of direct bandgaps. (e) The plane-wave interpolated GW-calculated VBM and CBM energies across the high-symmetry stackings in the moiré superlattice. All VBM are at $K$. CBM for AB1 is at $K$, while for AA' and AB2 are at $M$.

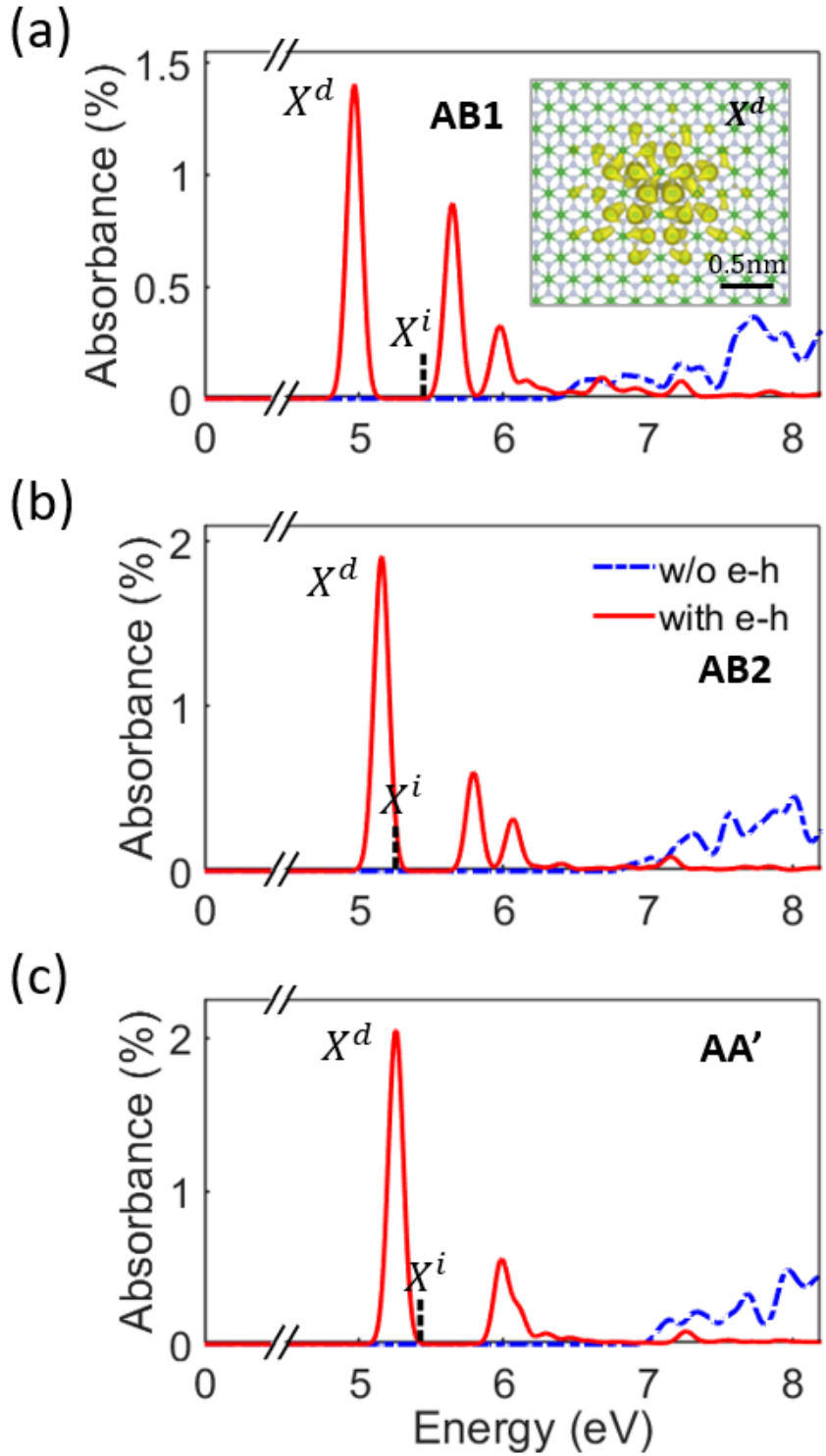

**Figure 3** Optical absorption spectra of (a) AA', (b) AB1 and (c) AB2 stacked bilayer *h*-BN without (blue dashed line) and with (solid red line) electron-hole interactions included. The direct $X^d$ and indirect $X^i$ states are labeled in the plot. The inset in (a) represents the real-space exciton wavefunction of $X^d$. The hole position is marked with a gray circle.

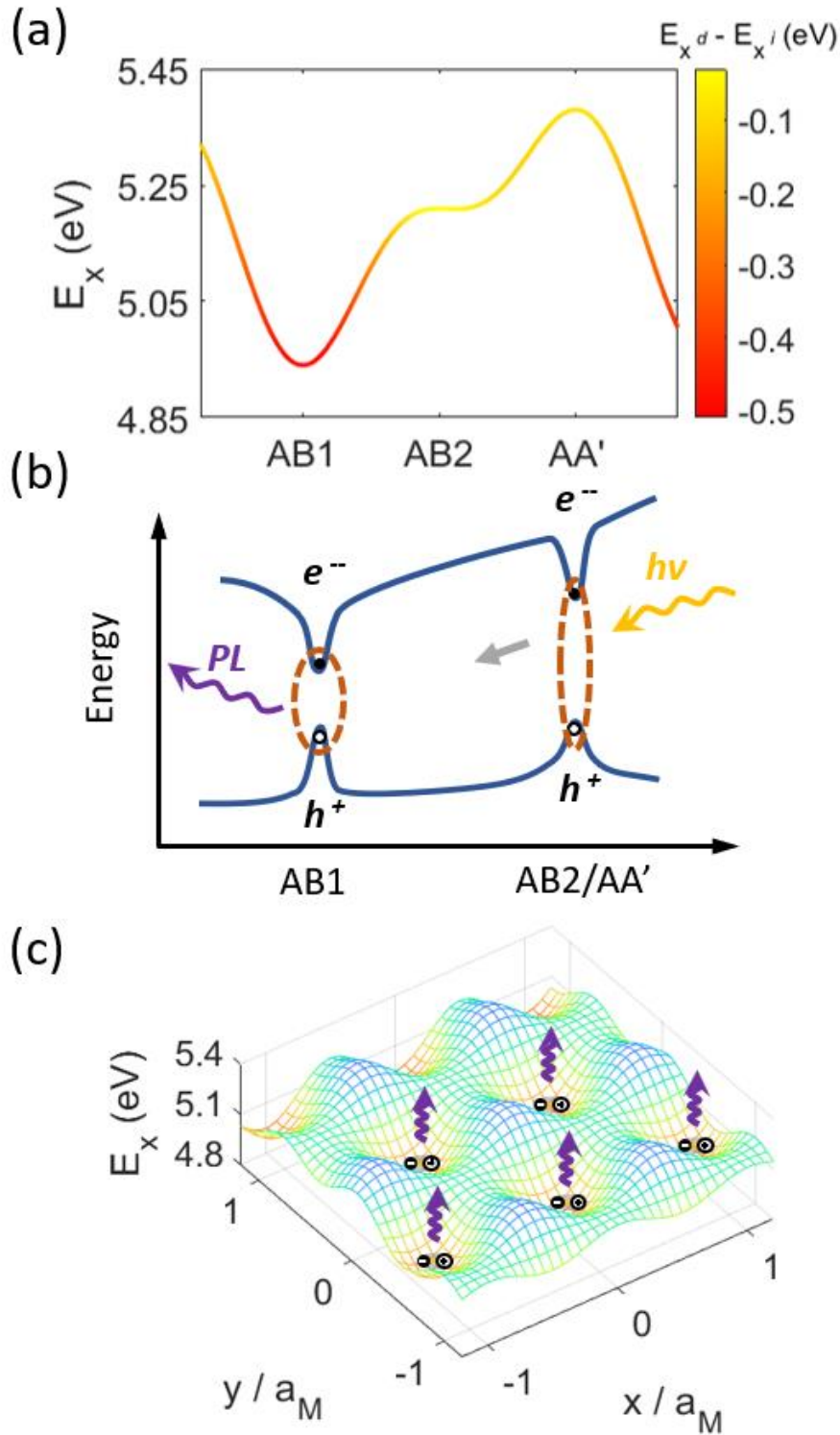

**Figure 4** (a) Exciton energy profile in the moiré superlattice. The color bar represents the energy difference between direct $X^d$ and indirect $X^i$ excitons at local stackings. (b) Schematic plot of the modified band edge landscape under photoexcitation due to the electron-hole binding effect. The large exciton binding results in exciton funneling to AB1. The grey arrow shows the motion of excited carriers under incident light. (c) Exciton energy profile in the moiré superlattice, in which the photoexcited excitons are accumulated around the minimum-excitonic-energy region round AB1.

**Table of Contents:**

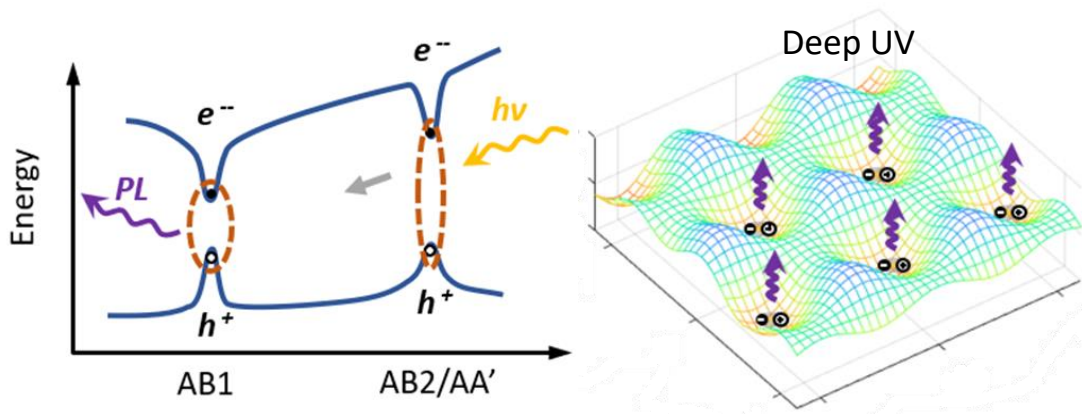